\documentclass[letterpaper,twocolumn,american]{revtex4-1}
\usepackage{lmodern}
\usepackage[T1]{fontenc}
\usepackage[latin9]{inputenc}
\usepackage{color}
\usepackage{float}
\usepackage{amsmath}
\usepackage{amssymb}
\usepackage{graphicx}

\makeatletter

\usepackage{slashed}
\definecolor{teal}{RGB}{52,90,138}
\definecolor{orange}{RGB}{52,90,138}

\makeatother

\usepackage{babel}
\begin{document}

\title{\noindent Topological invariants for the Haldane phase of interacting
SSH chains -- a functional RG approach}

\author{Björn Sbierski and Christoph Karrasch}

\affiliation{Dahlem Center for Complex Quantum Systems and Institut für Theoretische
Physik, Freie Universität Berlin, 14195, Berlin, Germany}

\date{\today}
\begin{abstract}
We present a functional renormalization group approach to interacting
topological Green function invariants with a focus on the nature of
transitions. The method is applied to chiral symmetric
fermion chains in the Mott limit that can be driven into a Haldane
phase. We explicitly show that the transition to this phase is accompanied
by a zero of the fermion Green function. Our results for the phase boundary are quantitatively benchmarked against DMRG data.
\end{abstract}
\maketitle

\section{Introduction}

The complete topological classification of non-interacting fermionic
insulators and superconductors was a milestone achievement in condensed
matter theory \cite{Kitaev2009,Ryu2010a,Qi2010a,Hasan2010}. The
result is conveniently summarized in the ten-fold way table which
lists the equivalence classes of Hamiltonians depending on spatial
dimension and the presence of time-reversal, particle-hole and chiral
symmetries. In terms of Bloch Hamiltonians $\mathcal{H}(\mathbf{k})$,
this amounts to investigating the topological properties of the map
$\mathbf{k}\rightarrow\mathcal{H}(\mathbf{k})$. Two $\mathcal{H}(\mathbf{k})$
are equivalent if they can be deformed into each other without breaking
the specified symmetries or closing the gap.

Soon after, efforts were directed to a generalization for interacting
systems, leading to the concept of symmetry protected topological
states (SPT) \cite{Pollmann2010,Fidkowski2010,Fidkowski2011,Turner2013,Wen2017}.
The (non-degenerate) ground states of two many-body
Hamiltonians are equivalent if they can be adiabatically connected
without breaking the defining symmetries (which is possible if and only if the Hamiltonians can be deformed into each other without closing of the many-body gap).
To date, the topological classification for fermionic interacting
system is not known completely, except in one spatial dimension.

Given a certain microscopic model, one would like to know its ground
state's equivalence class, usually as a function of the model parameter.
This is achieved in terms of topological invariants, which can be
formulated in various equivalent ways. In the noninteracting case, the invariant can be based on the eigenstates of Bloch
Hamiltonians $\mathcal{H}(\mathbf{k})$. Given a control parameter
in $\mathcal{H}$, it can be shown that the (integer valued) topological
invariant $\nu\left(\mathcal{H}\right)$ can only change at gapless
points where $\mathcal{H}(\mathbf{k})$ has zero eigenvalues at some
momentum $\mathbf{k}$ in the Brillouin zone: If two Bloch Hamiltonians feature different (the same) invariants, they cannot (can always) be deformed into each other without closing the gap.

In the interacting case, one can still consider the noninteracting expressions for the invariants if one replaces the Bloch Hamiltonian with the inverse single-particle retarded $T=0$ Green function at vanishing
frequency, $\mathcal{H}(\mathbf{k})\rightarrow-G^{-1}(i\omega=0,\mathbf{k})$
\cite{Volovik2003}. In the mathematical formulation of $\nu\left(G\right)$
to be detailed below, $G$ and $G^{-1}$ are used on equal footing
and correspondingly, $\nu$ can change at poles of $G$, where $G^{-1}=0$
for some momentum in the Brillouin zone, or at zeros with $G=0$ \cite{Gurarie2011}.
A pole is interpreted as a closing of the single-particle excitation gap whereas
a zero indicates a breakdown of the single-particle picture and is
ruled out in the noninteracting case (for bounded Hamiltonians). As
shown in Ref.~\cite{Manmana2012}, a zero can be both compatible with
a many-body gap closing (e.g., the spin gap closes while the charge
gap stays open) or with a unique, gapped ground state (no gap closing). It is therefore possible that two different noninteracting topological phases can be adiabatically deformed into each other when interactions are switched on but that the noninteracting invariant still changes, for a recent experimental proposal see Ref.~\cite{Yoshida2018}. Thus, a new classification becomes necessary in interacting systems. This is also reflected in the recently proposed many-body invariants of Refs. \cite{Shapourian2016,Shiozaki2017b}. 

\noindent 
\begin{figure}[b]
\noindent \begin{centering}
\includegraphics{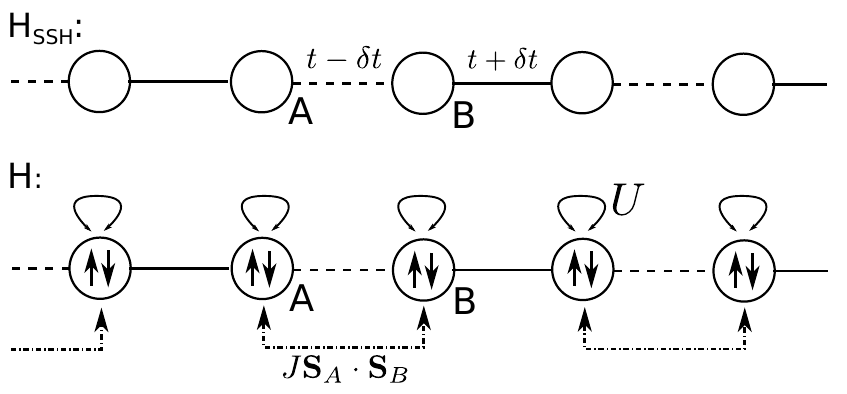}
\par\end{centering}
\caption{\label{fig:models}Model Hamiltonians used in this paper. The sublattices
are denoted by A and B, straight lines denote single-particle hopping
$t\pm\delta t$ and curved arrows denote Hubbard interactions $U$.
Spin-Spin interactions $J$ are denoted by straight arrows.}
\end{figure}

In the following, we focus on the evaluation of the Green function
invariant $\nu\left(G\right)$ with an emphasis on the nature of the
transition points. Considerable effort has been directed to one-dimensional
systems. In many cases of interest, the Green function can be calculated
analytically. For example, You et al.~used an unconventional perturbation
theory in the non-interacting part of the Hamiltonian to demonstrate that when a topological phase transition between two noninteracting
phases is gapped by interactions, the poles will be replaced by Green
function zeros \cite{You2014}. Moreover,
in Ref.~\cite{Manmana2012}, Green functions at transition points
were calculated analytically for several models at special points
in parameter space. In the general case, however, a numerical evaluation of the Green function is
required. Previous studies \cite{Manmana2012,Yoshida2014} employed the
density matrix renormalization group (DMRG) \cite{White1992} to compute the Green function
winding number. Although the DMRG and its underlying matrix-product
state formulation is very well suited to determine one-dimensional
topological phases via entanglement properties \cite{Pollmann2010}, it has severe shortcomings
when it comes to calculating Green function winding numbers: In order
to compute $\nu(G)$, the Green function is required at zero frequency
$G(i\omega=0,\mathbf{k})$ in the thermodynamic limit, which is generally difficult for the DMRG \cite{Schollwock2005,Schollwoeck2011}. One can, e.g., use a real-time algorithm to generate Green functions via a Fourier transform; the accessible timescales, however, are limited by the entanglement growth, and instead of $i\omega=0$, one can determine $G$ reliably only at finite frequencies. Since the invariant
is quantized, this should not affect the results, except in the vicinity
of points in parameter space where $\nu$ changes. However, as we
discussed above, a precise assessment of the type of singularity occurring
at these points (Green function zero or pole) is essential.

In this paper, we propose the fermionic functional renormalization
group (fRG) \cite{Kopietz2010,Metzner2012} as an alternative method
to numerically evaluate Green function invariants. We show that the
fRG, set up in a Matsubara formulation and momentum space, is capable
of evaluating Green functions at $i\omega=0$ and easily tells poles
from zeros. We put an emphasis on transition points and show in detail
how the zeros of $G$ are understood in the framework of the self-energy \cite{Honerkamp2003}.
Although bulk-boundary correspondence
is often discussed in the context of topological systems, in the following we limit ourselves to the bulk
perspective (though fRG can also be applied to finite systems). While
the fRG can be set up in arbitrary dimension, for a concrete example, we focus on
one-dimensional systems with both charge conservation and many-body
chiral symmetry, i.e., interacting variants of the Su-Schrieffer-Heeger
(SSH) chain \cite{Su1979}. In this case, $\nu\left(G\right)$ takes
the form of a winding number and the classification is $\mathbb{Z}$
for noninteracting and $\mathbb{Z}_{4}$ for interacting systems \cite{Fidkowski2010,Fidkowski2011,Turner2011}.
A common criticism of the fRG method is its perturbative character.
Although the Green functions calculated within our fRG truncation
scheme below are guaranteed to be correct to second order in the interaction
only, the fRG results contain partial resummation of diagrams to infinite
order. For our models, we show that we can capture Mott physics both
qualitatively and quantitatively with reasonable accuracy. 

The rest of the paper is structured as follows. In Sec. \ref{sec:Model-Hamiltonians}
we present the model Hamiltonian and discuss its topological phase
diagram qualitatively. In Sec. \ref{sec:Green-function-topological}
we define the appropriate Green function winding number and show how
it is computed from the self-energy found by fRG. In Sec. \ref{sec:fRG}
we explain the fRG approach. The numerical results are presented in
Sec. \ref{sec:Results} along with a comparison to DMRG and we conclude
in Sec. \ref{sec:Conclusion}.

\section{Model Hamiltonian\label{sec:Model-Hamiltonians} }

\noindent We start by defining the Hamiltonian $H$ that we will employ
in the following (see Fig. \ref{fig:models} for a sketch), closely following Refs. \cite{Manmana2012,Yoshida2014}:
\begin{eqnarray}
H_{\mathrm{SSH}} & \!= & \!-\!\!\sum_{j}[\left(t\!-\!\delta t\right)\!c_{j,A}^{\dagger}c_{j,B}\label{HSSH}\\
 &  & \!+\!\left(t\!+\!\delta t\right)\!c_{j,B}^{\dagger}c_{j+1,A}\!+\!h.c.]\nonumber \\
H & \!= & \!H_{\mathrm{SSH},\uparrow}\!+H_{\mathrm{SSH},\downarrow}\label{H}\\
 & + & U\!\!\sum_{j;s=A,B}\!\left(n_{j,s,\uparrow}\!-\!\frac{1}{2}\right)\!\left(n_{j,s,\downarrow}\!-\!\frac{1}{2}\right)\nonumber \\
 & + & J\sum_{j}\mathbf{S}_{j,A}\cdot\mathbf{S}_{j,B}.\nonumber 
\end{eqnarray}
We denote fermion creators/annihilators on site $j$ of an infinite
one-dimensional lattice by $c_{j,s}^{(\dagger)}$, with each unit cell
split into sublattices $s=\{A,B\}$. The lattice constant is set to
unity and we work at half-filling throughout. The Hamiltonian
$H_{\mathrm{SSH}}$ is the SSH model which features a single spinless fermion
per lattice site and hoppings alternating between $t+\delta t$ and
$t-\delta t$. We then generalize to spinful fermions $c_{j,s,\sigma}^{(\dagger)}$
in $H$ and introduce on-site Hubbard interactions $U$ as well as an intra-unit cell spin-spin exchange interaction $J$. Here,
$\mathbf{S}=(S^{x},S^{y},S^{z})^{\mathrm{T}}$ denotes the spin operator
where $S^{i}=\frac{1}{2}\sum_{\sigma,\sigma^{\prime}}c_{\sigma}^{\text{\ensuremath{\dagger}}}\sigma_{\sigma\sigma^{\prime}}^{i}c_{\sigma^{\prime}}$
for $i=x,y,z$ (site/sublattice indices suppressed).

The model Hamiltonians $H_{\mathrm{SSH}}$ and $H$ are invariant under
time-reversal, particle-hole, and chiral symmetry; the single-particle version of $H_{\mathrm{SSH}}$ falls into the class BDI of the noninteracting Altland-Zirnbauer classification \cite{Altland1997}. The formulation of the topological invariant rests on chiral symmetry, it takes the form \cite{Gurarie2011}
\begin{equation}
H=U_{C}^{\dagger}H^{\star}U_{C},\label{chiral sym}
\end{equation}
where the star denotes complex conjugation not affecting fermionic
operators. The action of $U_{C}$ on fermion creation and anihilation operators is defined as
\begin{align}
U_{C}^{\dagger}c_{\alpha}U_{C} & =\sum_{\beta}c_{\beta}^{\dagger}\left[\tau_{z}\right]_{\beta\alpha},\\
U_{C}^{\dagger}c_{\alpha}^{\dagger}U_{C} & =\sum_{\beta}\left[\tau_{z}\right]_{\alpha\beta}c_{\beta},
\end{align}
where $\alpha,\beta$ contain the single-particle
indices as appropriate for the different models discussed above and
$\tau_{z}$ is the third Pauli matrix in sublattice space. Note that
the site- and spin-labels (if present) are not modified. Before
we investigate the restrictions on the single-particle ground state
Green function arising from chiral symmetry and formulate the winding
number, we qualitatively discuss the physics of $H_{\mathrm{SSH}}$
and $H$, following Ref. \cite{Manmana2012}.

The SSH chain $H_{\mathrm{SSH}}$ is gapped for $\delta t\ne0$. For
the special point $\delta t=t$, there is no hopping between A and
B sublattice sites of the same unit cell while there is dimerization
between B and A sublattice site across the unit cell. If we would terminate
the chain at an A site, we would have a single-particle edge state, thus
$\delta t>0$ corresponds to the topological, $\delta t<0$ to the
trivial phase. At the transition point $\delta t=0$, the Green function
has a pole. Now consider $H$ with $J=0$, i.e., a spinful SSH chain featuring a Hubbard interaction
$U>0$. At $\delta t=0$, the system is a Mott insulator at half filling
with low energy spin-1/2 degrees of freedom coupled anti-ferromagnetically
with strength $\sim t^{2}/U>0$ \cite{Lieb1968,Manmana2012,Barbiero2018}. This half-integer
spin chain has a charge gap but gapless spin excitations. For $\delta t\neq0$,
these spin couplings alternate in strength, the spins pair up,
form singlets, and we obtain a spin gap. In conclusion, the addition
of the Hubbard term $U>0$ does not modify the topological phase diagram
from the case $U=0$; however, the transition in $\nu(G)$ at $\delta t=0$
is now accompanied by a zero of the Green function reflective of the collective
nature of the gapless spin excitation when expressed in terms of fermion
operators. Since the non-interacting Hamiltonian
$H_{\mathrm{SSH}}$ vanishes for $\delta t=0$ and $k=\pi$, the appearance
of a Green function zero can also be derived using perturbation theory
as in Ref. \cite{You2014}. 

We now consider $\delta t<0$ (which gaps $H_{\mathrm{SSH}}$)
and switch on a finite spin-spin exchange interaction $J$ in $H$, which we choose to be negative (ferromagnetic). For $|J|\gg t^{2}/U$, it leads to the formation of effective
spin-1 objects in each unit cell which are coupled anti-ferromagnetically.
This state is known to be in the Haldane phase\cite{Manmana2012},
which is gapped and topological with spin-1/2 edge excitations, again
hinting towards a closing of a spin gap and corresponding Green function
zero at the transition point. In the following, we keep $J<0$ constant
but increase $U$ to tune the transition from a trivial phase at $|J|\ll t^{2}/U$ to the Haldane
phase for $|J|\gg t^{2}/U$. It is the central goal of this paper to show that the fermionic
fRG is capable of detecting the Haldane phase via the Green function
winding number and unambiguously identifies the zero at the transition.
We note that in Ref. \cite{Anfuso2007} it was demonstrated that the
Haldane phase built from spin-1/2 fermions can be adiabatically connected
to a trivial phase even without gap closing, but this required the
breaking of chiral symmetry.

\section{Green function winding number\label{sec:Green-function-topological}}
We start the discussion of topological invariants from the non-interacting
SSH model $H_{\mathrm{SSH}}$, Eq. (\ref{HSSH}). After a spatial
Fourier-transform, $c_{j,s}=\int_{-\pi}^{\pi}\frac{dk}{2\pi}c_{k,s}e^{ikj}$,
we obtain
\begin{equation}
H_{\mathrm{SSH}}=\int_{-\pi}^{\pi}\frac{dk}{2\pi}\,c_{k}^{\dagger}\mathcal{H}_{\mathrm{SSH}}(k)c_{k}, 
\end{equation}
where $c_{k}=(c_{k,A},\,c_{k,B})^{\mathrm{T}}$. The corresponding Bloch Hamiltonian
reads 
\begin{equation}
\mathcal{H}_{\mathrm{SSH}}(k)=\left(\begin{array}{cc}
0 & h_{\mathrm{SSH}}(k)\\
h_{\mathrm{SSH}}^{\dagger}(k) & 0
\end{array}\right),\label{HI_Bloch}
\end{equation}
with $h_{\mathrm{SSH}}(k)=-\left(t+\delta t\right)e^{-ik}-\left(t-\delta t\right)$.
Topological invariants for non-interacting insulators with chiral symmetry in odd dimensions are winding numbers of $\mathbb{Z}$-type.
In one dimension, the invariant can be expressed as \cite{Volovik2003}
\begin{align}
\nu_{\mathrm{SSH}} & =\int_{-\pi}^{\pi}\frac{dk}{4\pi i}\,\mathrm{tr}\left[\tau_{z}\mathcal{H}_{\mathrm{SSH}}^{-1}(k)\partial_{k}\mathcal{H}_{\mathrm{SSH}}(k)\right]\label{eq:nu_H}\\
 & =\int_{-\pi}^{\pi}\frac{dk}{2\pi i}\,\partial_{k}\mathrm{log}\,h_{\mathrm{SSH}}^{\dagger}(k),\nonumber 
\end{align}
counting how often $h_{\mathrm{SSH}}(k)$ winds around
the origin of the complex plan, $\nu_{\mathrm{SSH}}\in\mathbb{Z}$. The winding is trivial (zero)
for $\delta t<0$ and nontrivial for $\delta t>0$. The off-diagonal
form of Eq. (\ref{HI_Bloch}), and thus the existence of the winding
number $\nu_{\mathrm{SSH}}$, is a consequence of the chiral symmetry,
Eq. (\ref{chiral sym}), which enforces
\begin{equation}
\mathcal{H}_{\mathrm{SSH}}(k)=-\tau_{z}\mathcal{H}_{\mathrm{SSH}}\left(k\right)\tau_{z}.\label{eq:chiral_H}
\end{equation}

We now generalize the definition of the winding number to arbitrary
chiral, translational invariant and possibly interacting
systems featuring a a gap of single-particle excitations. The central object of the following
discussion is the imaginary frequency Green function (at zero temperature).
It is defined as a Fourier transform of the imaginary time Green function,
\begin{equation}
G(i\omega)=\int_{0}^{\infty}d\tau\,e^{i\omega\tau}G(\tau),~ G_{\alpha\beta}(\tau)=-\left\langle T_{\tau}c_{\alpha}\left(\tau\right)c_{\beta}^{\dagger}\right\rangle,
\end{equation}
where $c_{\alpha}\left(\tau\right)=e^{H\tau}c_{\alpha}\,e^{-H\tau}$, $T_{\tau}$ denotes time-ordering, and $\alpha,\beta$ are single-particle
multiindices $\alpha,\beta$. After a spatial Fourier transform the
Green function $G(i\omega,k)$ is diagonal in the crystal momentum $k$.
It can be shown \cite{Gurarie2011} that under a general chiral symmetry
(which can always be represented by $\tau_{z}$ in some basis), $G$ transforms as
\begin{equation}
G^{-1}(i\omega,k)=-\tau_{z}G^{-1}\left(-i\omega,k\right)\tau_{z}.\label{G_chiral_sym}
\end{equation}
Analogous statements are true for other symmetries
\cite{Gurarie2011}. Based on Eq.~(\ref{G_chiral_sym}), winding numbers 
were defined, first using integration both over $\omega$ and $k$ \cite{Volovik2003,Gurarie2011}.
However, it was shown in Ref. \onlinecite{Wang2012a} that, given a non-singular $G(i\omega,k)$, all topological information
is contained in the Green function at $i\omega=0$, which is a well-defined limit for a gapped
system. In this case, Eq. (\ref{G_chiral_sym})
implies the form
\begin{equation}
-G^{-1}(i\omega=0,k)=\left(\begin{array}{cc}
0 & h\left(k\right)\\
h^{\dagger}\left(k\right) & 0
\end{array}\right),\label{eq:Gm1}
\end{equation}
with the depicted matrix structure in sublattice space and $h\left(k\right)$
a matrix in the remaining degrees of freedom. The Green function winding
number then reads \cite{Manmana2012}
\begin{align}
\nu & =\int_{-\pi}^{\pi}\frac{dk}{4\pi i}\,\mathrm{tr}\left[\tau_{z}G(0,k)\partial_{k}G^{-1}(0,k)\right]\label{nu_G}\\
 & =\int_{-\pi}^{\pi}\frac{dk}{2\pi i}\,\mathrm{tr}\left[\partial_{k}\mathrm{log}\,h^\dagger(k)\right],\nonumber 
\end{align}
counting the complex plane winding of the eigenvalues of $h(k)$ around
the origin. Mathematically, the robustness of $\nu$ can be formulated
as follows: Assume that the system {[}and $h(k)${]} depends on some
external parameter $\xi$, then the winding number is invariant under small
changes of $\xi$, as one can see from $\partial_{\xi}\nu=0$ \cite{Manmana2012}.
For noninteracting systems with Bloch Hamiltonian $\mathcal{H}(k)$, from the relation $G(i\omega,k)=1/\left[i\omega-\mathcal{H}(k)\right]$,
we have $\,-G^{-1}(i\omega=0,k)=\mathcal{H}(k)$ so that Eq. (\ref{nu_G})
specializes to Eq. (\ref{eq:nu_H}) with $h\rightarrow h_{\mathrm{SSH},\uparrow}+h_{\mathrm{SSH},\downarrow}$.

Based on Eq.~(\ref{nu_G}), it is evident that besides poles {[}vanishing
eigenvalues of $G^{-1}(0,k)${]} also zeros {[}vanishing eigenvalues
of $G(0,k)${]} can cause a change of the winding number $\nu$ \cite{Gurarie2011}.
Poles are familiar from the non-interacting case and indicate a zero-energy single-particle excitation. The presence of zeros indicates
a complete loss of single particle coherence and is an inherent many-body
phenomenon. By passing through zeros, the winding number can change
without a gap closing. This is the mechanism that causes the collapse
of free classification of topological fermion phases in the presence
of interactions. Alternatively, the zero can occur
along with a many-body gap closing; in this case, it signals a topological
phase transition.

Given a generic interacting fermion system, it is thus desirable to
devise a numerical method to (i) compute the winding number $\nu$
(or Chern-number, as appropriate) in the case that $G(0,k)$ is non-singular,
and (ii) classify the nature of the points in parameter space where
$\nu$ changes, i.e., tell poles from zeros. It is our goal to show
how the fRG can be used for both purposes.

The central object obtained from the fRG is the single-particle
self-energy $\Sigma(i\omega,k)$. The Green function is then given
by
\begin{equation}
G\left(i\omega,k\right)=\frac{1}{i\omega-\mathcal{H}_{0}(k)-\Sigma\left(i\omega,k\right)},\label{eq:G_selfEnergy}
\end{equation}
where $\mathcal{H}_{0}(k)$ is the non-interacting Bloch Hamiltonian.
The presence of a zero is tied to a vanishing quasiparticle weight
for some $k$ in the Brillouin zone \cite{Honerkamp2003,BruusBook},
\begin{equation}
Z(k)\equiv\left(1-\partial_{\omega}\mathrm{Im}\Sigma(i\omega,k)|_{\omega=0}\right)^{-1}.\label{eq:Z(k)}
\end{equation}
This becomes apparent if one rewrites 
\begin{equation}
G\left(i\omega\simeq0,k\right)=\frac{Z(k)}{i\omega-\mathcal{H}_{\mathrm{top}}(k)},
\end{equation}
with
\begin{align}
\mathcal{H}_{\mathrm{top}}(k) & =Z(k)\left[\mathcal{H}_{0}(k)+\Sigma\left(i\omega=0,k\right)\right].
\end{align}
If $Z(k)$ is finite, the Green function winding number can be obtained from $\mathcal{H}_{\mathrm{top}}(k)$,
which is off-diagonal as in Eq. (\ref{eq:Gm1}). A vanishing eigenvalue
of $\mathcal{H}_{\mathrm{top}}(k)$ at finite $Z(k)$ indicates the
presence of a Green function pole. 

\section{Functional RG\label{sec:fRG}}

The functional renormalization group method is an implementation
of the RG idea on the basis of many-body vertex functions, see Refs.
\cite{Kopietz2010,Metzner2012} for general introductions. The idea
amounts to using an infrared cutoff $\Lambda$ in the bare Matsubara
Green function $G_{0}\left(i\omega,k\right)=\left(i\omega-\mathcal{H}_{0}(k)\right)^{-1}$,
here we choose a frequency cutoff $G_{0}^{\Lambda}\left(i\omega,k\right)=\Theta\left(|\omega|-\Lambda\right)G_{0}\left(i\omega,k\right).$
Then, the $\Lambda$-dependence carries over to all vertex functions,
the simplest of which is the self energy $\Sigma^{\Lambda}\left(i\omega,k\right)$
appearing in the full Green function, see Eq. (\ref{eq:G_selfEnergy}).
In the limit $\Lambda=\infty$, the dynamics of the system is frozen
and the vertex functions are trivial. The fRG flow equations
are an infinite set of coupled differential equations that describe
the change of the vertex functions with $\Lambda$. The solution of
these flow equations at $\Lambda=0$ (where the cutoff vanishes) yields
exact vertex functions of the physical problem. In practice, truncation
of the infinite hierarchy of flow equations is required and the resulting
vertex functions approximate the exact ones with an agreement of at
least order $O(v^{n})$ where $v$ is a proxy for the interaction
strength in the Hamiltonian (i.e., $U$ or $J$ in Eq. (\ref{H})),
and $n=1,2,3,...$ depending on the level of the truncation. Note
that unlike perturbation theory, the fRG contains an infinite re-summation
of Feynman diagrams. We have recently developed a $k$-space fRG approach which is
correct to order $O(v^{2})$ for one-dimensional, translationally invariant
fermion systems in equilibrium, see Ref. \onlinecite{Sbierski2017}. There, we have applied the fRG to a Luttinger liquid with good agreement to alternative exact methods. We refer the reader to Ref. \onlinecite{Sbierski2017} for further discussion of the method.

For the self-energy, the flow equation reads \cite{Sbierski2017} 
\begin{eqnarray}
\partial_{\Lambda}\Sigma_{\alpha^{\prime}\alpha}^{\Lambda}\left(i\omega,k\right) & \!= & \!-\!\int_{-\pi}^{\pi}\!\frac{d\bar{k}}{2\pi}\!\int_{-\infty}^{+\infty}\!\frac{d\bar{\omega}}{2\pi}\sum_{\beta,\beta^{\prime}}S_{\beta\beta^{\prime}}^{\Lambda}\left(i\bar{\omega},\bar{k}\right)\nonumber \\
 & \times & V_{\beta^{\prime}\alpha^{\prime};\beta\alpha}^{\Lambda}\left(\begin{array}{ccc}
i\bar{\omega},\bar{k}; & i\omega,k; & i\bar{\omega},\bar{k}\end{array}\right),\label{eq:flowS}
\end{eqnarray}
where $S^{\Lambda}\left(i\omega,k\right)$ is the single-scale propagator,
$\ensuremath{S^{\Lambda}=G^{\Lambda}\left(\partial_{\Lambda}\left[G_{0}^{\Lambda}\right]^{-1}\right)G^{\Lambda}}$,
and $V^{\Lambda}$ the 2-particle vertex where frequency and momentum conservation
has been used to eliminate the fourth argument. Initially, $V^\Lambda$ is frequency independent,
$V^{\Lambda=\infty}\sim U+J$. In a first-order trunctation ($n=1$), the flow
of $V^{\Lambda}$ (which is itself of order $v$) would be neglected by setting $V^{\Lambda}\rightarrow V^{\Lambda=\infty}$ in Eq. (\ref{eq:flowS}).
Evidently, $\Sigma^{\Lambda}\left(i\omega,k\right)$ then turns out
to be frequency independent, and consequently $Z(k)=1$ as is apparent
from Eq. (\ref{eq:Z(k)}). Thus, the truncation to order $n=2$
with a flowing and frequency-dependent 2-particle vertex $V^{\Lambda}$
is mandatory for our purpose. The full flow equations, including a static but fully momentum dependent feedback for $V^{\Lambda}$
are lengthy and are given in Eqs. (34)-(36) of Ref. \onlinecite{Sbierski2017}.

\section{Results\label{sec:Results}}

\noindent 
\begin{figure}[t]
\noindent \begin{centering}
\includegraphics{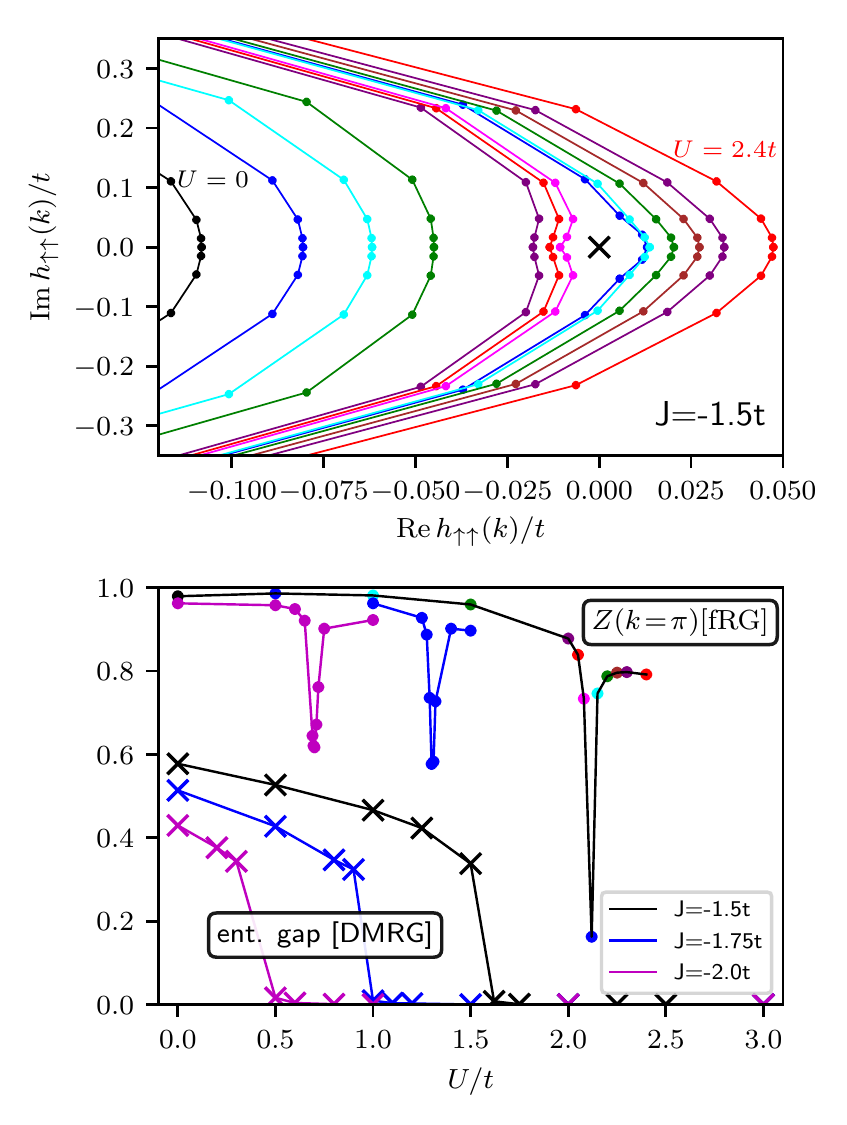}
\par\end{centering}
\caption{\label{fig:results}Top panel: fRG results for the Green function
winding of the chiral fermion chain $H$ for fixed $\delta t=-t/4$,
$J=-1.5t$ with increasing $U$, driving the transition from a trivial
to the Haldane phase at $U_{c,\mathrm{fRG}}\simeq2.1t$. The finite
value of $|h_{\uparrow\uparrow}(k)|$ for all $k$ and $U$ reveals
the presence of a gap of single-particle excitations across the transition. Bottom
panel: The vanishing of the quasiparticle weight $Z(k\!=\!\pi)$ (colored points on black line) confirms the
presence of a Green function zero at the transition. The comparison
with the more precise DMRG result $U_{c,\mathrm{DMRG}}\simeq1.6$ based
on the entanglement gap indicates a slight overestimation of $U_{c}$
by the fRG. Qualitatively similar results hold for $J=-1.75$ (blue symbols) and $J=-2$ (magenta symbols) for which $U_c$ decreases.}
\end{figure}

We now proceed to present the fRG results for the winding number and the quasiparticle
weight for the chiral fermion chain $H$ in Eq. (\ref{H}). Except at the critical point mentioned at the following, the Green function $G(i\omega,k)$ is found to be regular for all $i\omega$ and $k$ and thus the simplified expression (\ref{nu_G}) can be applied. 
As a phase
diagram (based on the DMRG entanglement spectrum) can be found
in the appendix of Ref. \cite{Yoshida2014}, we focus on a single
line in parameter space. We let $\delta t=-t/4$,
$J=-1.5t$ and increase the Hubbard interaction $U$ to drive the
transition from a trivial to a Haldane insulator once $|J|\gg t^{2}/U$.
Note that the non-interacting part of $H$ is gapped and convergence
issues of the fRG as encountered for $\delta t=0$ in Ref. \onlinecite{Sbierski2017}
are absent. Due to $S_{z}$ conservation and rotation symmetry, the
off-diagonal blocks of $-G^{-1}(i\omega=0,k)$ are of the form
\begin{equation}
h(k)=\left(\begin{array}{cc}
h_{\uparrow\uparrow}(k) & 0\\
0 & h_{\downarrow\downarrow}(k)
\end{array}\right)
\end{equation}
with $h_{\uparrow\uparrow}=h_{\downarrow\downarrow}$. In Fig. \ref{fig:results},
the top panel depicts the complex value of $h_{\uparrow\uparrow}(k)$
for increasing $U$ in the vicinity of $k=\pi$ (identified by $\mathrm{Im}\left[h_{\uparrow\uparrow}(k=\pi)\right]=0$).
The origin of the complex plane is denoted by a black cross. The phase winding of $h_{\uparrow\uparrow}(k)$ is trivial (leading to $\nu=0$)
for $U<U_{c,\mathrm{fRG}}$ and non-trivial (winding once around the origin, $|\nu|=2$ due to spin)
for $U>U_{c,\mathrm{fRG}}$ with $U_{c,\mathrm{fRG}}\simeq2.1t$.
For all $U$ and $k$, the magnitude of $|h_{\uparrow\uparrow}|$ is
larger than a constant, signaling the presence of gapped single-particle excitations
(no pole) throughout the transition.

The lower panel depicts the behavior of $Z(k\!=\!\pi)$ (thin black line) which
sharply drops in the vicinity of the transition and confirms the presence
of a Green function zero at the transition. The lowest value found
is below 0.2 for $U=2.12t$. To gauge the quantitative
reliability of the fRG results, we have calculated the phase boundary
using the entanglement gap from DMRG [imaginary time evolution with a bond dimension of $\chi=320$, see Refs. \onlinecite{Schollwoeck2011} and \onlinecite{Karrasch2016}]. The data is shown as black crosses and signals the
transition at a critical value of $U_{c,\mathrm{DMRG}}\simeq1.6$,
slightly smaller than the fRG value. Qualitatively similar results hold for different values of $J$, see blue and magenta symbols for $J=-1.75$ and $J=-2$, respectively.

\section{Conclusion\label{sec:Conclusion}}

We presented the fermionic fRG method as a valuable tool to study
SPT phases in terms of Green function winding numbers. Our emphasis
was on the nature of the transition between different phases. We
explicitly showed for a topological Mott insulator chain how zeros
of the Green function can be unambiguously identified from the self-energy. Although the phase diagram itself can be determined using more accurate methods such as the DMRG, obtaining the Green function in the limit $i\omega\rightarrow0$ is a difficult task, and the fRG offers complementary, qualitative information. It would be
interesting to apply the fRG to higher dimensional SPTs where accurate reference methods are sparse and the nature of the transition is
less obvious. We remark that similar applications have been put forward
in the ``Hierachy of correlations'' approach of Ref. \onlinecite{Gomez-Leon2016}.

\section*{Acknowledgement}

We acknowledge useful discussions with Lisa Markhof, Achim Rosch, Robert Peters
and Carolin Wille. Numerical computations were done on the HPC cluster
of Fachbereich Physik at FU Berlin. Financial support was granted
by the Deutsche Forschungsgemeinschaft through the Emmy Noether program
(KA 3360/2-1) and the CRC/Transregio 183 (Project
B01).

\bibliographystyle{apsrev4-1}
\bibliography{/home/bjoern/Desktop/Research/library}

\end{document}